\documentclass{aa}
\usepackage{times}
\usepackage{graphics}
\usepackage{epsf}
\begin{document}
%\topmargin0.0cm
%
%%%\thesaurus{06(03.20.1; 08.03.4; 08.09.2: WR\,118; 08.13.2; 08.16.4; 13.09.6)}
%
\title{
Speckle interferometry and radiative transfer 
modelling of the Wolf-Rayet star \object{WR\,118}
}
\author{
B.\ Yudin\inst{1}\and
Y.\ Balega\inst{2}\and
T.\ Bl\"ocker\inst{3}\and
K.-H.\ Hofmann\inst{3}\and
D.\ Schertl\inst{3}\and
G. Weigelt\inst{3}
}
\institute{
Sternberg Astronomical Institute, Universitetskii pr. 13,
119899 Moscow, Russia \and
Special Astrophysical Observatory, Nizhnij Arkhyz, Zelenchuk region,
Karachai--Cherkesia, 35147, Russia (balega@sao.ru) \and
Max--Planck--Institut f\"ur Radioastronomie, Auf dem H\"ugel 69,
D--53121 Bonn, Germany \\
(bloecker@mpifr-bonn.mpg.de, hofmann@mpifr-bonn.mpg.de, 
schertl@mpifr-bonn.mpg.de, weigelt@mpifr-bonn.mpg.de)
}
\offprints{B.\,Yudin (yudin@sai.msu.ru)}
%\mail{yudin@sai.msu.ru}
%

%%%%% \date{this version: \today}
%%%%% \date{submitted: \today}
%%%%\date{revised version: \today}
%%%%\date{final version edp: \today}
%%%\date{received ~~~~ / accepted ~~~~}
\date{Received date /  accepted date}
\titlerunning{ 
Speckle interferometry and radiative transfer 
modelling of \object{WR\,118}
}
\authorrunning{B.\ Yudin et al.}
\abstract{
\object{WR\,118}
is a highly evolved Wolf-Rayet star of the WC10 subtype surrounded 
by a permanent dust shell absorbing and re-emitting in the infrared 
a considerable fraction of the stellar luminosity. 
We present the first diffraction-limited 2.13\,$\mu$m speckle interferometric 
observations of \object{WR\,118} with 
73 mas resolution. The speckle interferograms were obtained with the
6\,m telescope at the 
Special Astrophysical Observatory.
The two-dimensional visibility function of the object does not show any 
significant deviation from circular symmetry.
The visibility curve declines towards the 
diffraction cut-off frequency to $\sim0.66$ and  can be approximated  by
a linear function. 
%This means that the dust shell has a great albedo ($>0.3$) at 2.13\,$\mu$m
%and contains therefore a significant amount (in mass fraction) of grains 
%larger than 0.3\,$\mu$m.
Radiative transfer calculations have been carried out to model the spectral energy 
distribution, given in the range of 0.5-25\,$\mu$m, and our 2.13\,$\mu$m visibility function, 
assuming spherical symmetry of the dust shell. Both can be fitted with a
model containing double-sized grains (``small'' and ``large'') with the radii of 
a = 0.05\,$\mu$m and 0.38$\mu$m, and a mass fraction of the large grains greater than 65\%. 
Alternatively, a  good match can be obtained with 
the grain size distribution function $n(a) \sim a^{-3}$, with $a$ 
ranging between 0.005$\mu$m and 0.6$\mu$m. At the inner boundary of the modelled dust shell 
(angular diameter $\Theta_{\rm in} = (17 \pm 1)$\,mas),
the temperature of the smallest  grains and the dust shell density are 
1750\,K $\pm$ 100\,K and $(1 \pm 0.2) \cdot 10^{-19}$\,g/cm$^{3}$, respectively. 
The dust formation rate is found to be $(1.3 \pm 0.5) \cdot 10^{-7}$ M$_{\odot}$/yr, 
assuming $V_{\rm wind} = 1200$\,km/s. 
%%%The distance estimates are in the range 1.5-2.2 kpc.
%
%
\keywords{
Techniques: image processing ---
Circumstellar matter ---
Stars: individual: WR\,118 ---
Stars: mass--loss ---
Stars: Wolf-Rayet ---
Infrared: stars
}
}
\maketitle
\section{Introduction}
The Wolf-Rayet star \object{WR\,118}  
(= \object{CRL\,2179} = \object{IRAS 18289-1001})
in the catalog of WR stars of van der Hucht (2001) 
belongs to the latest subtypes of the carbon-rich sequence (WC).
It is classified as WC9-10 (Allen et al.\ 1977, Massey \& Conti 1983). 
Its very prominent IR excess is attributed to the thermal 
emission of a warm carbon dust shell. No remarkable variations in the dust emission 
have been registered over the last two decades (Williams et al.\ 1987, 
van der Hucht et al.\ 1996) and it has been designated as a ``persistent'' 
dust-maker (Williams \& van der Hucht 2000). 
While the dust grains are momentum-coupled to the fast stellar wind, 
fresh dust grains should be steadily condensing at a definite distance from the star, 
i.e.\ the dust formation rate is generally constant on the time scale of several decades.

Among the known galactic WR stars, 
\object{WR\,118} is one of the optically faintest objects, but
one of the brightest ones in the infrared regime.
This is naturally explained by its heavy interstellar reddening and the obscuration 
by its conspicuous dust shell. 
A strong interstellar 9.7\,$\mu$m silicate feature is 
observed in its spectrum with an optical depth of $\tau_{9.7{\rm m}} \approx 0.7$ 
(Roche \& Aitken 1984, van der Hucht et al.\ 1996). 
Using the relation $A_{V}/\tau_{9.7\mu{\rm m}} \leq 21.5$ (Schutte et al.\ 1998), 
an extinction of $A_{V} \leq 15$ can be derived. 
Allen et al.\ (1977) and Williams et al.\ (1987) list $A_{V} \sim 15$ and 
$\sim 12.8$, respectively.
In the optical region, the brightness of WR\,118 is only roughly known.

From spectrophotometry, Massey \& Conti (1983) estimate  the $v$ narrow-band  magnitude to 
be approximately 22, and Cohen \& Vogel (1978) list a $V$ broad-band magnitude of  $\sim 20$.   
Williams et al.\ (1987) modelled the IR spectral energy distribution (SED) of the 
WC stars, finding that they could be well fitted with an optically thin shell 
consisting of small ($\sim 0.01$\,$\mu$m) amorphous carbon grains 
with a density distribution of $\rho \propto r^{-2}$. 
The absence of the silicon carbide feature at 11.3\,$\mu$m in the spectra of WR stars 
exclude SiC to be a constituent of their dust shells. 
In this paper we present diffraction-limited 73 mas 
speckle interferometric observations of the dust shell of \object{WR\,118}. 
Radiative transfer  calculations have been performed to model both the 
SED and the 2.13\,$\mu$m visibility.
\section{Observations and data reduction}
\begin{figure}[bhtp]
\epsfxsize=8.8cm 
%%%%%\mbox{\epsffile[-60 -50 270 270]{\LHOME/tex/wr118/edp/plots3/WR118-9909-K.nord_p2.0.ima.2.eps}}\\
\mbox{\epsffile[-60 -50 270 270]{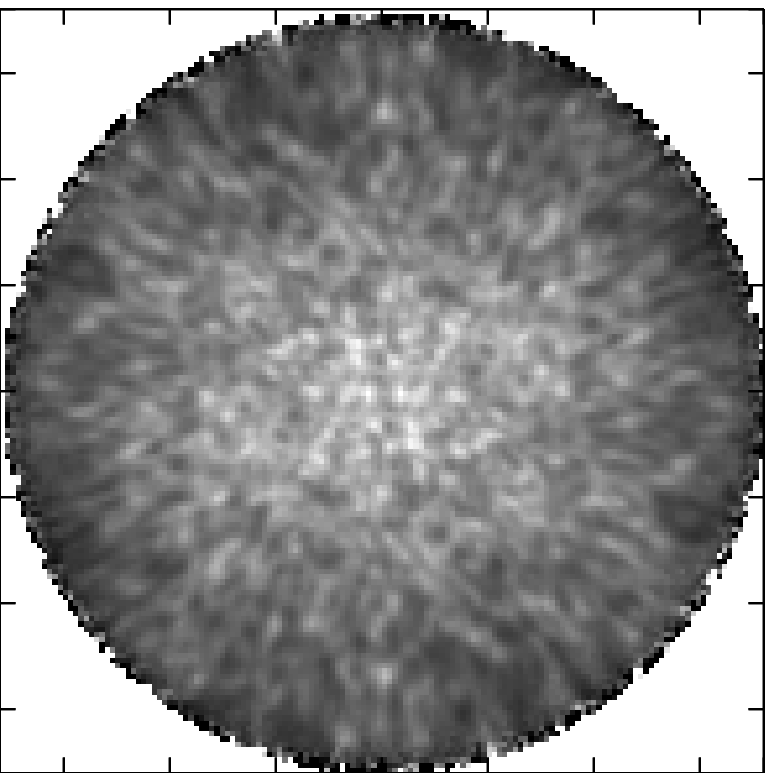}}\\
\vspace*{5mm}
\epsfxsize=8cm
%%%%%\mbox{\epsffile{\LHOME/tex/wr118/edp/plots3/WR118-9909-K.vis5.5.eps}} 
\mbox{\epsffile{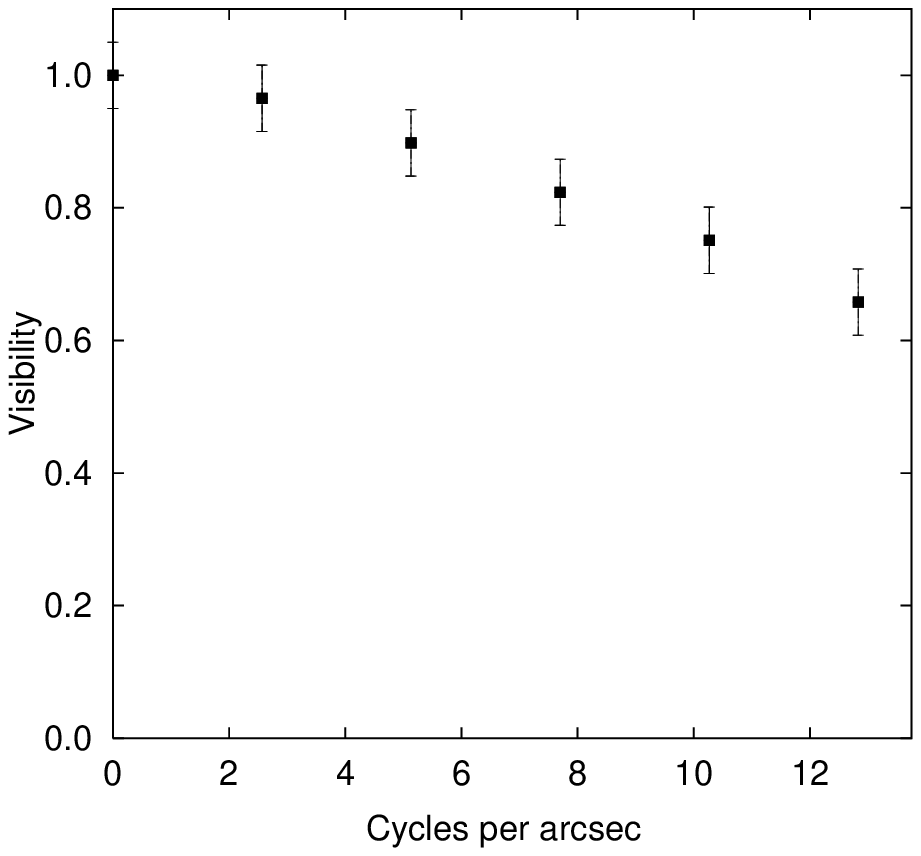}} 
\caption{
Two-dimensional 2.13\,$\mu$m visibility (top) 
and azimuthally averaged 2.13\,$\mu$m visibility 
of \object{WR\,118} (bottom) shown up to the diffraction limit.
%The light central structure shows that the central object is
%surrounded by a dust shell
}
\label{Fima} 
\end{figure}

The \object{WR\,118} speckle interferograms were obtained with the Russian 6\,m 
telescope at the Special Astrophysical Observatory on September 27, 1999. The speckle data were 
recorded with HAWAII speckle camera (HgCdTe array, 256$^{2}$ pixels, sensitivity from 1 to 
2.5\,$\mu$m, frame rate 2 frames/s) through an interference filter with a
central wavelength of 
2.13\,$\mu$m and a bandwidth of 0.21\,$\mu$m. Speckle interferograms of the unresolved star 
\object{HIP\,87540} 
were taken for the determination of the atmospheric speckle transfer function. The 
observational parameters were as follows: exposure time/frame 20 ms; number of frames 
210 (130 of \object{WR 118} and 80 of {HIP 87540}; 
2.13\,$\mu$m seeing (FWHM) 1\farcs4;
%%%2.13\,$\mu$m seeing (FWHM) 0\farcs9 (with image motion  1\farcs4);
field of view 5\farcs1 x 5\farcs1; pixel size 26.4 mas. 
The resolution was 73\,mas.
%%%(Weigelt 1977, Lohmann et al. 1983, Hofmann \& Weigelt 1986, Weigelt 1991). 
The visibility function
%%%(modulus of the object Fourier transform) 
(modulus of the Fourier transform of the object intensity) 
was determined with the speckle interferometry method (Labeyrie 1970).
Figure~\ref{Fima} shows the two-dimensional 2.13\,$\mu$m visibility 
and the azimuthally averaged 2.13\,$\mu$m visibility of WR\,118 up
to the diffraction cut-off frequency.
There is no evidence for deviation from circular symmetry. 
The visibility curve can be approximated by a linear function and dropped to ~0.66 at 
the diffraction cut-off frequency.

%%%%%%%%%%%%%%%%%%%%%%%%%%%%%%%%%%%%%%%%%%%%%%%%%%%%%%%%%%%%%%%%%%%%%%%%%%%%%
%%%% Images (top) and contour plots(bottom) K
%%%%%%%%%%%%%%%%%%%%%%%%%%%%%%%%%%%%%%%%%%%%%%%%%%%%%%%%%%%%%%%%%%%%%%%%%%%%%

\section{The dust shell model}
Using the quasi-diffusion method of Leung (1975), as implemented in the
CSDUST3 code (Egan et al.\ 1988),
we have constructed spherically symmetric radiative transfer models 
to match both the observed SED and the 2.13\,$\mu$m visibility. 
The input parameters of the model are: (i) the spectral shape and bolometric 
luminosity of the star ($f(\lambda) L_{\rm bol}$),
after passing the shell, $f(\lambda)$ $L_{\rm bol}$ transforms to 
$f^{\prime}(\lambda) L_{\rm bol}$;
(ii) the inner shell radius ($r_{\rm in}$); 
(iii) the relative thickness of the shell, i.e.\ the ratio of outer to inner 
shell radius ($Y_{\rm out} = r_{\rm out}/r_{\rm in}$); 
(iv) the density distribution $\rho(r)$; 
(v) the chemical composition and grain-size distribution; 
(vi) the total optical depth at a given reference wavelength.

The  input parameter of the radiative transfer model, which relates the 
stellar bolometric luminosity and the inner shell radius, is the incident flux at the 
inner boundary of the dust shell: 
$F_{\rm in} = L_{\rm bol}/(4\pi r_{\rm in}^{2})$. 
It is based on the assumption that the 
temperature of the dust grains is controlled by the radiation field only.
%%and that the dust grains'  optical properties do not depend on surroundings. 
This means that if we change $L_{\rm bol}$ but 
keep $F_{\rm in} = const$ (i.e.\ change $r_{\rm in} \propto \sqrt{L_{\rm bol}}$, resp.) 
we obtain (i) the same SED; (ii) the 
same normalized surface brightness distribution as a function of %% (scaled to) 
$b/r_{\rm in}$   ($b/r_{\rm in}$: impact parameter); and
(iii) the same visibility curve as  function of %%(scaled to) 
$q\Theta_{\rm in}$ ($q$: spatial frequency; 
$\Theta_{\rm in}$: angular diameter of the dust shell's inner boundary, 
$b/2r_{\rm in}=\Theta/\Theta_{\rm in}$). 
For more details of self-similarity and scaling behaviour of infrared 
emission from radiatively heated dust,
see also Ivezic \& Elitzur (1997) and references 
therein. For the sake of clarity, we have 
incorporated $L_{\rm bol}$ and $r_{\rm in}$ separately  in the CSDUST3 code
instead of using $F_{\rm in}$ only. 
During the SED fitting procedure,
we kept $L_{\rm bol}$ constant., i.e. changed only
$r_{\rm in}$.

Keeping in mind that the distance is defined by the relation 
$d = \sqrt{f^{\prime}(\lambda) L_{\rm bol}/4\pi F_{\rm obs}(\lambda)}$
($F_{\rm obs}(\lambda)$: observed (dereddened) spectral energy flux),
i.e.\ $d \propto \sqrt{L_{\rm bol}}$, 
it can be shown that the angular surface brightness distribution of the 
model, in particular $\Theta_{\rm in} = 2 r_{\rm in}/d$, does not depend on 
$L_{\rm bol}$. In our calculations  $\Theta_{\rm in}$  is determined by  
comparing the modelled and observed $K$-band ($\lambda = 2.2$\,$\mu$m) flux. 
We note that, if the observed SED is properly fitted, we get the same 
value of $\Theta_{\rm in}$ conducting the flux normalization  at any
other wavelength.
In principle, the bolometric flux can be used as well,
but for WR 118 this appears to be unfavorable due to the lack of
optical and UV data.

\begin{table}[htbp] 
  \caption{
Model dust-shell parameters. The models M\,1, M\,2, M\,1.1 and M\,2.1
refer to 
double-sized grains with $a=0.05$\,$\mu$m and  $a=0.38$\,$\mu$m, the model
M\,3 to a grain size distribution of $n(a) \propto a^{-3}$ with 
$0.005\,\mu$m$ \leq a \leq 0.6\,\mu$m. $\tau_{V}$ is the optical depth at 
0.55\,$\mu$m;  $r_{\rm in}$, $T_{\rm in}$ and  $\Theta_{\rm in}$ are
the radius, the temperature (refering to the smallest grains)
and the angular diameter of the inner dust shell rim,  
$n_{0.05}$ and  $m_{0.05}$ are the abundances 
of 0.05\,$\mu$m grains by number and mass, resp., and
$A_{K}$ is the 2.2\,$\mu$m dust-shell albedo.
%The interstellar extinction adopted amounted to 
%$A_{V}=12.8$ (models  M\,1, M\,2, M\,3) and 
%$A_{V}=14.0$ (models  M\,1.1, M\,2.1), resp.
  }
  \label{Tlumindep} 
  \begin{tabular}{lcccccccc}
    \hline\noalign{\smallskip}
     model& $\tau_{V}$  & $r_{\rm in}$ & $T_{\rm in}  $   & $n_{0.05} $  
          & $m_{0.05} $ & $A_{K}$      & $\Theta_{\rm in}$ \\
          &             & ($R_{\ast}$) &   (K)            & (\%) 
          &  (\%)       &              &    (mas)         \\
    \noalign{\smallskip}
    \hline\noalign{\smallskip}       
    M\,1   & 0.88 & 179 & 1670 & 99.55 & 34 & 0.36 & 17.8 \\
    M\,2   & 0.70 & 140 & 1820 & 98.83 & 17 & 0.40 & 16.3 \\
    M\,3   & 0.70 & 152 & 1760 &       &    & 0.36 & 17.5 \\
    M\,1.1 & 0.73 & 159 & 1740 & 99.55 & 34 & 0.36 & 17.4 \\
    M\,2.1 & 0.58 & 125 & 1900 & 98.83 & 17 & 0.40 & 16.0 \\
    \noalign{\smallskip}
    \hline\noalign{\smallskip} 
  \end{tabular}
\end{table}
%%%%%%%%%%%%%%%%%%%%%%%%%%%%%%%%%%%%%%%%%%%%%%%%%%%%%%%%%%%%%%%%%%%%%%%%%%%%%%%%%%%%%%%%%%%%
\begin{table*}[bhtp] 
  \caption{
Predicted \object{WR\,118} parameters:
color $V-J$; dereddened and reddenend visual magnitudes $V_{0}$ and $V$;
distance $d$, bolometric luminosity $L_{\rm bol}$; radius  $r_{\rm in}$
and density $\rho_{\rm in}$ at inner dust shell rim;
and dust mass-loss rate  $\dot{M}_{\rm dust}$.
%The observed $V$ magnitude is $\sim 20$ (Cohen \& Vogel 1978). For 
%$A_{V}=12.8$ (models  M\,1, M\,2, M\,3) and 
%$A_{V}=14.0$ (models  M\,1.1,
%M\,2.1
%) the observed $J_{0}$ magnitude is 4.65 and 4.31, resp.
%(Allen et al.\ 1977, Williams et al.\ 1987).
  }
  \label{Tlumdep} 
  \begin{tabular}{lcccccccc}
    \hline\noalign{\smallskip}
     model& $V-J$               & $V_{0}$            & $V$             & $d$  
          & $L_{\rm bol}$       & $r_{\rm in}$       & $\rho_{\rm in}$ & $\dot{M}_{\rm dust}$ \\
          &                     & (mag)              &   (mag)         & (kpc) 
          & ($10^{4}L_{\odot}$) & ($10^{3}R_{\odot}$)& ($10^{-19}$g/cm$^{3}$)  &($10^{-7}M_{\odot}$/yr) \\
    \noalign{\smallskip}
    \hline\noalign{\smallskip}       
    M\,1   & 2.22 & 6.87 & 19.7 & 2.3 & 6.7 & 4.2 & 0.9 & 1.8\\
    M\,2   & 1.68 & 6.33 & 19.1 & 1.8 & 5.6 & 3.0 & 1.2 & 1.3\\
    M\,3   & 1.67 & 6.32 & 19.1 & 1.8 & 5.6 & 3.2 & 0.8 & 1.0\\
    M\,1.1 & 2.15 & 6.46 & 20.5 & 1.9 & 5.5 & 3.4 & 0.9 & 1.2\\
    M\,2.1 & 1.62 & 5.93 & 19.9 & 1.5 & 4.7 & 2.4 & 1.2 & 0.8\\
    \noalign{\smallskip}
    \hline\noalign{\smallskip} 
  \end{tabular}
\end{table*}

In our model calculations, we firstly assumed a bolometric luminosity of    
$L_{\rm bol} = 4 \cdot 10^{4} L_{\odot}$.
Luminosity-independent model parameters are given in Tab.~\ref{Tlumindep}. 
Removing the reddening from the observed $J$ magnitude of \object{WR\,118} 
and taking the model $V-J$ color,
we obtain the (unreddened) model $V_{0}$ magnitude of \object{WR\,118}. 

Adopting a  
WC9-10 absolute magnitude of $M_{V} = -4.9$ (Williams et al.\ 1987), 
$d$ and $L_{\rm bol}$ can be estimated.  Taking this 
new value of $L_{\rm bol}$ (see Tab.~\ref{Tlumdep}) we recalculated the 
luminosity-dependent model parameters such as $r_{\rm in}$, 
the rate of dust formation at the inner boundary of the dust shell
($\dot{M}_{\rm dust}$), and the density of the dust shell at the inner
boundary ($\rho_{\rm in}$) using the relations 
$r_{\rm in} \propto \sqrt{L_{\rm bol}}$, 
$\dot{M}_{\rm dust} \propto \sqrt{L_{\rm bol}}$, and 
$\rho_{\rm in} \propto 1/ \sqrt{L_{\rm bol}}$.
Concerning the  dust formation rate, an 
outflow velocity of  $V_{\rm wind} = 1200$\,km/s was adopted (Williams et al. 1987). 
The model predictions for all these luminosity dependent parameters of \object{WR\,118} 
are given in Tab.~\ref{Tlumdep}. 

For the SED of the star, $f(\lambda)$, we took basically a black body of 
19000 K (van der Hucht et al. 1986). We corrected it in the region of        
0.5-1.25\,$\mu$m to have $f(\lambda) \propto \lambda^{-3.0}$
(Williams et al. 1998) corresponding to $(V-J) \approx 0.0$.  In 
the infrared, corrections for the continuum radiation
of the ionized gas shell have been made to obtain 
$(J-K) \approx 0.25$ and $f(\lambda) \propto \lambda^{-2.7}$
longwards of 10\,$\mu$m
(Williams et al.\ 1987). It should be noted that these corrections have only minor effects 
on the model SED due to a very strong IR excess caused by the emission of
dust grains.

The density distribution of the dust shell is given by 
$\rho \propto r^{-2}$, i.e by a uniform outflow model.
The outer shell boundary was chosen
to be $Y_{\rm out}=10^{3}$. Larger values of  $Y_{\rm out}$ affect the SED 
only in the far-infrared, i.e.\ far beyond the IRAS 25\,$\mu$m
photometric band, 
which defines the long-wavelength tail in this study. 
Concerning the  optical dust-grain constants,
we considered  the amorphous carbon grains (cel1000) of   
J\"ager et al.\ (1998). The reference wavelength of the optical depth was chosen to be
0.55\,$\mu$m, and the model SED is normalized to the observed one in the $J$ band. 
%$\alpha$ being close to 2. At the long 
%wavelength side the SED of WR 118 is restricted to the 25 ?m IRAS photometric band. 
%Thus we can determine only the low limit on value of the dust shell thickness, i.e. fix 
%Yout, whose increasing it does not effect on the SED at the region of 25 ?m. We take Yout 
%= 103, that is ?3 times greater them its low limit. 
%we take the total optical depth at ? = 
%0.55 ?m (?V). The model SED we normalize to the observed one at J spectral band.
%Concerning the  optical dust-grain constants

\begin{figure}[bhtp]
\begin{centering}
\epsfxsize=8cm
%%%\mbox{\epsffile[145 275 404 530]{\LHOME/tex/wr118/edp/figures/Fig_2.ps}} 
\mbox{\epsffile[145 275 404 530]{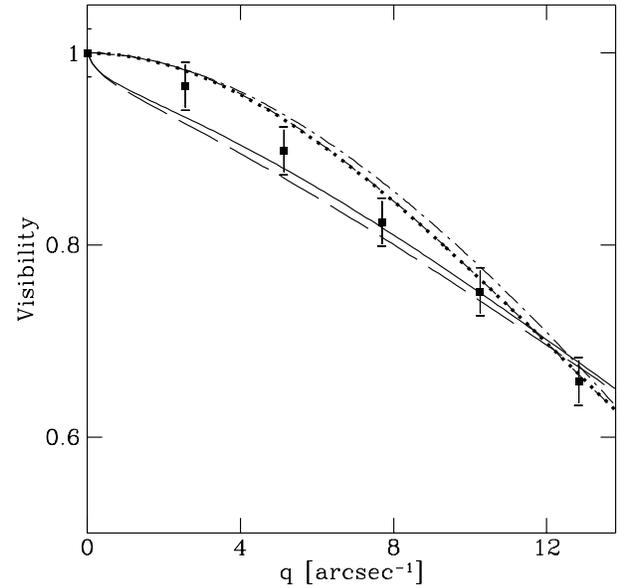}} 
\end{centering}
\caption{
2.13\,$\mu$m model visibility function of \object{WR 118} for different
optical depths and grain sizes for a dust-shell with $r_{\rm in}=200 R_{\ast}$
together the observed visibility. 
The dotted, short-dashed, dot-dashed, solid and long-dashed lines refer to
$\tau_{V}=0.6$, $a=0.01\mu$m;
$\tau_{V}=0.6$, $a=0.10\mu$m;
$\tau_{V}=4.0$, $a=0.10\mu$m;
$\tau_{V}=4.0$, $a=0.38\mu$m; and
$\tau_{V}=0.6$, $a=0.38\mu$m, resp.
}
\label{Fvisigrain} 
\end{figure}

Figure~\ref{Fvisigrain} shows the 2.13\,$\mu$m model visibility  
for different optical depths (0.6 and 4) and grain sizes
(single-sized grains with 0.01\,$\mu$m, 0.1\,$\mu$m and 0.38\,$\mu$m)
for a dust shell with $r_{\rm in}/R_{\ast}=200$ in comparison with the observed visibility.
Only for the model with the largest grains, the visibility 
curve approaches to a linear function, i.e. its shape approaches 
the one of the observed visibility of \object{WR\,118}. 
%%%%%%%%%%%%%%%%%%%%%%%%%%%%%%%%%%%%%%%%%%%%%%%%%%%%%%%%%%%%%%%%%%%%%%%%%%%%%
%%%% K band visibility
%%%%%%%%%%%%%%%%%%%%%%%%%%%%%%%%%%%%%%%%%%%%%%%%%%%%%%%%%%%%%%%%%%%%%%%%%%%%%
  
The transformation of the shape of the  2.13\,$\mu$m visibility curve from parabolic to 
rectilinear is connected with the transformation of a mostly thermally emitting dust shell
to a highly scattering one when the albedo of dust grains is getting large. 
The 2.13\,$\mu$m  albedo of the J\"ager et al. (1998) carbon grains with a=0.1\,$\mu$m is 
0.04, i.e.\  the dust shell's radiation is dominated by direct thermal emission
at this wavelength. 
The albedo of the corresponding 0.38\,$\mu$m  grains is 0.4, i.e. the dust shell 
scatters the IR light very efficiently.  Increasing the optical depth of an
``emitting'' and a ``scattering'' dust shell 
results into a slight increase and decrease of the 
visibility's curvature, respectively (Fig.~\ref{Fvisigrain}).
 
Keeping in mind the very hostile environment for dust condensation in the shells 
of WR stars, it appears to be obvious to assume that the dust grains are rather small 
(Williams et al.\ 1987). However, 
such models will not be able to reproduce the shape of the 
visibility curve of \object{WR\,118}. 
On the other hand, it is clear that the dust shell cannot consist 
entirely of large grains, which grow up from their small progenitors. In our 
double-sized grains models, we considered 
the combination of small grains ($a = 0.05$\,$\mu$m) 
and large grains ($a = 0.38$\,$\mu$m). 

The temperature of the large grains is $\sim 1.5$ times lower
than that of the small ones. 
Thus, if one increases the abundance of the large grains, $r_{\rm in}$ has to be decreased
to preserve the SED fits in the near-infrared domain. 
Although this even worsens the condition of dust-grain condensation 
in the corresponding models, the observed shape of the visibility 
curve clearly requires to include large grains in the dust shell of \object{WR\,118}.

The decrease of the small grains' size does not affect  the 
SED nor the visibility substantially, i.e. the grains with
$a \le 0.05$\,$\mu$m 
behave similarly, in this respect, in the optical and IR domain. 
The increase of the relative abundance of the small grains  
mainly reflects in an increase of the emergent radiation's reddening
in the optical 
region, and a decrease of the dust shell's albedo in the infrared. 
The latter leads to an essential change of the visibility's curvature in the ``wrong'' direction. 
Both these points define the abundance of small grains in the dust shell. 
However, in the case of \object{WR\,118}, we 
have no accurate estimations of its brightness in the 
optical wavelength range, and thus cannot 
determine how much reddening we need.  
%%It prevents also to put in the model AV as its free parameter. 
Calculations were performed adopting $A_{V} = 12.8$ and 14, which leads to 
$J_{0} \approx 4.65$ and 4.31 (Allen et al. 1977, Williams et al. 1987), resp.

The increase of the large grains' size does not affect the  
SED and the 2.13\,$\mu$m visibility very much, either. 
However, the size of the large grains 
cannot be decreased significantly  because the 2.13\,$\mu$m
albedo of the J\"ager et al. (1998) grains with
$a=0.25$\,$\mu$m is $\sim 0.27$ 
being already too low to allow a proper fit of the visibility.
Besides double-sized grains models, we also 
calculated models with a power-law grain-size 
distribution of $n(a) \propto a^{-\gamma}$
($ 0.005\,\mu$m$ \leq a \leq 0.6\,\mu$m) 
to investigate which exponent $\gamma$ is most appropriate. 
For all these models,
$T_{\rm in}$ is calculated for the smallest dust grains with 
$ a = 0.005\,\mu$m.
This temperature exceeded that of the $ a = 0.05\,\mu$m grains
by $\sim 5$\%.

\section{Results of calculations and discussion} \label{Ssum}
\begin{figure}[tbp]
\begin{center}
\epsfxsize=6cm
%%%\mbox{\epsffile[145 160 404 693]{\LHOME/tex/wr118/edp/figures/Fig_3.ps}} 
\mbox{\epsffile[145 160 404 693]{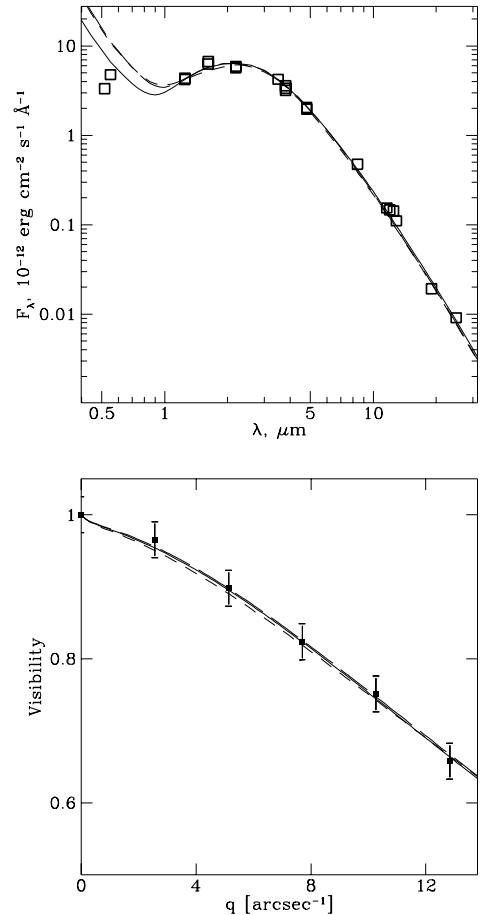}} 
\end{center}
\caption{
Model SED corrected for interstellar extinction of $A_{V}=12.8$ and 2.13\,$\mu$m visibility function 
of \object{WR 118} for the models M1 (solid lines), M2 (short-dashed lines), and M3 (long-dashed lines)
as given in Tabs.~\ref{Tlumindep}-\ref{Tlumdep}. The squares refer to the observations (see text). 
}
\label{Fsedvisi_12_8} 
\end{figure}

\begin{figure}[thbp]
\begin{center}
\epsfxsize=6cm
%%%\mbox{\epsffile[145 160 404 693]{\LHOME/tex/wr118/edp/figures/Fig_4.ps}} 
\mbox{\epsffile[145 160 404 693]{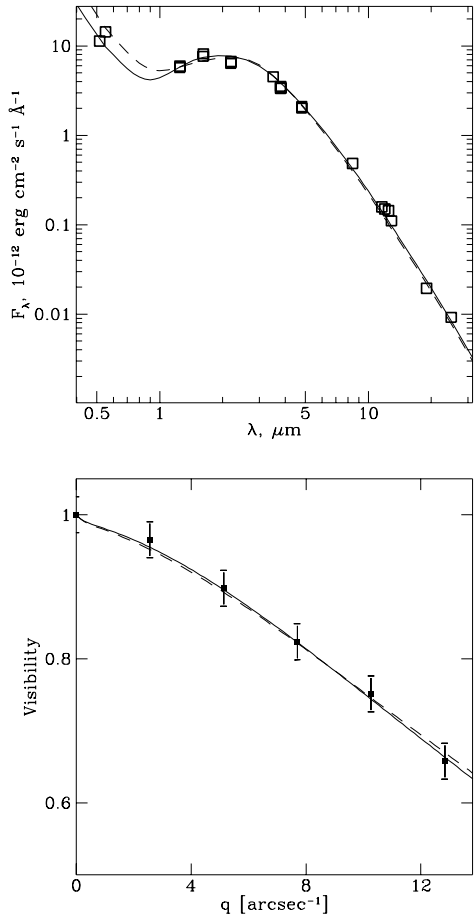}} 
\end{center}
\caption{
Model SED corrected for interstellar extinction of $A_{V}=14.0$ and 2.13\,$\mu$m visibility function 
of \object{WR 118} for the models M1 (solid lines), M2 (short-dashed lines), and M3 (long-dashed lines)
as given in Tabs.~\ref{Tlumindep}-\ref{Tlumdep}. The squares refer to the observations (see text). 
}
\label{Fsedvisi_14_0} 
\end{figure}

Figures~\ref{Fsedvisi_12_8}-\ref{Fsedvisi_14_0}
show the observed 2.13\,$\mu$m visibility of WR 118 and its SED corrected 
for interstellar extinction of $A_{V}= 12.8$ and 14, respectively. The latter includes the data 
of Massey \& Conti (1983) ($V$ band), Cohen \& Vogel (1978) ($V$ band), Allen et al. (1977) 
($JHKL$ bands), Williams et al. (1987) $JHKL\prime M$ [8.4], [11.6], [12.5], [12.9], and [19] 
bands) and color-corrected  IRAS photometry (IRAS Science Team 1988) ([12] and [25] 
bands).
Figures ~\ref{Fsedvisi_12_8} and \ref{Fsedvisi_14_0}
also show  the 2.13\,$\mu$m visibility and SED of those 5 models, which 
proved to be best suited for the dust shell of WR\,118 in the framework of spherical symmetry. Their 
parameters are given in Tab.~\ref{Tlumindep} and \ref{Tlumdep}. 
For comparison, Williams et al.\ (1987) have found 
that the dust shell of \object{WR\,118} have the following parameters: 
$r_{\rm in} = 260 R_{\ast}$, % = 3.8 \cdot 10^{3} R_{\odot}$ ($d$=???????),  
$Y_{\rm out}= 30$,
$T_{\rm in} =1410$\,K, $\rho_{\rm in} = 4.9 \cdot 10{-19}$\,g/cm-3 and 
$\dot{M}_{\rm dust} = 8.2 \cdot 10^{-7}$\,M$_{\odot}$/yr.  
%%%%%%%%%%%%%%%%%%%%%%%%%%%%%%%%%%%%%%%%%%%%%%%%%%%%%%%%%%%%%%%%%%%%%%%%%%%%%
%%%% MODELS: SED and  K band visibility AV=12.8
%%%%%%%%%%%%%%%%%%%%%%%%%%%%%%%%%%%%%%%%%%%%%%%%%%%%%%%%%%%%%%%%%%%%%%%%%%%%%
%%%%%%%%%%%%%%%%%%%%%%%%%%%%%%%%%%%%%%%%%%%%%%%%%%%%%%%%%%%%%%%%%%%%%%%%%%%%%
%%%% MODELS: SED and  K band visibility AV=14.0
%%%%%%%%%%%%%%%%%%%%%%%%%%%%%%%%%%%%%%%%%%%%%%%%%%%%%%%%%%%%%%%%%%%%%%%%%%%%%

The models M1, M2, and  M3 refer to an interstellar extinction of
$A_{V}=12.8$, and the models
M1.1 and M2.1 to  $A_{V}=14.0$. M3 is the model with a
power law grain-size distribution with 
$\gamma = 3$. The other models refer to double-sized grains. 
In the latter case the relative abundance of 0.05\,$\mu$m grains 
in number ($n_{0.05}$) and mass ($m_{0.05}$)
are given in Tab.~\ref{Tlumindep},  
which also contains the optical depth of the dust
envelope $\tau_{V}$ and its albedo at 2.13\,$\mu$m  ($A_{K}$).

Regarding the models with the double-sized grains
($a=0.05\,\mu$m and $a=0.38\,\mu$m),
the relative concentration of the 0.05\,$\mu$m grains is  subject to some
uncertainty, 
since the small grains determine the reddening, which is not precisely known. 
A more accurate determination of this parameter demands a more precise estimation of 
the optical brightness of \object{WR\,118} at several wavelengths.
However, the grain density at the inner boundary of the dust shell and the dust formation 
rate do not depend strongly on the adopted $A_{\rm V}$ and relative grain concentration. 
Thus, the estimations given in Tab.~\ref{Tlumdep} can be considered as reliable 
in the framework of a spherically symmetric dust shell model.

Furthermore, it turned out that a match of the SED at the region of 25\,$\mu$m  
can be improved if the exponent in the power law of the dust density distribution 
is $\alpha = 2.1$ instead of 2. It remains open if this deviation from the 
uniform outflow is significant with respect to the observational and theoretical 
uncertainties. 
%%%That probably means that the grains were slightly destroyed while moving out with the 
%%%stellar wind. 
The same effect as increasing $\alpha$ can be 
reached by decreasing $Y_{\rm out}$ to 300. This is the lower limit for the dust thickness 
above which the 25\,$\mu$m flux is not affected any longer. However, 
such a sharp truncation 
of the outer parts of the dust shell looks less natural than, e.g., 
a gradual change of the mass-loss rate.  
%%%%%%than gradual sputtering of the dust grains in these layers. 
%The apparent lack of IR light variations on the time scale of 
%$300 r_{\rm in}/V_{\rm wind} < 25\,{\rm yr}$ 
%demonstrates that the stellar wind of WR 118 does not 
%change remarkably on such time interval in agreement 
%with the obtained value of $\alpha$ exponent.

The M1 and M3 models have almost the same albedo $A_{K}$, and the M2 and M3 models
have almost the same circumstellar (cst) reddening ($A_{V} - A_{J}$)/$A_{V}|_{\rm cst}$. 
It defines the similarity of the M1 and M3
visibility curves and the similarity of corresponding SEDs (Fig. 1). The mass 
fraction of the grains with $a \geq 0.3 \mu$m  in model M3 is $\sim 50$\%.  

\begin{table}[tbhp] 
  \caption{
Model dust-shell parameters for grain-size distributions with
$n(a) \propto a^{-\gamma}$ and $0.005\,\mu$m$ \leq a \leq 0.6\,\mu$m.
$\tau_{V}$ is the optical depth at 
0.55\,$\mu$m;  $r_{\rm in}$, $T_{\rm in}$ and  $\Theta_{\rm in}$ are
the radius, the temperature (refering to the smallest grains)
and the angular diameter of the inner dust-shell rim, 
and $A_{K}$ is the 2.2\,$\mu$m dust-shell albedo.
  }
  \label{Tgraindist} 
  \begin{tabular}{lcccccc}
    \hline\noalign{\smallskip}
     model& $\tau_{V}$  & $r_{\rm in}$ & $T_{\rm in}  $   & $\gamma $  
          & $A_{K}$     & $\Theta_{\rm in}$  \\
          &             & ($R_{\ast}$) &   (K)            &  
          &             &    (mas)        \\
    \noalign{\smallskip}
    \hline\noalign{\smallskip}       
    M\,4   & 0.87 & 197 & 1610 & 3.5 & 0.29 & 19.2 \\
    M\,3   & 0.70 & 152 & 1760 & 3.0 & 0.36 & 17.5 \\
    M\,5   & 0.56 & 124 & 1900 & 2.5 & 0.38 & 17.1 \\
    \noalign{\smallskip}
    \hline\noalign{\smallskip} 
  \end{tabular}
\end{table}
%%%%%%%%%%%%%%%%%%%%%%%%%%%%%%%%%%%%%%%%%%%%%%%%%%%%%%%%%%%%%%%%%%%%%%%%%%%%%%%%%%%%%%%%%%%%
Increasing the exponent $\gamma$ in the power-law grain-size distribution
from 3.0 (M3) to 3.5 (M4) leads to a decrease of the albedo
$A_{K}$ from 0.36 (Tab.~\ref{Tgraindist}) to 0.29, 
and to a significant increase of the visibility curvature,
and thus to a worse match of the observations.
The strong dependence of the visibility curve's shape on the albedo 
is confirmed by model calculations with double-sized grains.  
On the other hand, the decrease of $\gamma$ to 2.5 does not change the albedo $A_{K}$ much (to 
only $\sim 0.38$) and correspondingly the visibility curve either (model M5).  However, then
the model $V-J$ color decreases to 1.20 (1.67 for M3) in contradiction to the observations.
%%%that come to the rise of the predicted optical brightness of WR 118 
%%%which will exceed remarkably the observed one. 
Moreover,
the increase of the relative concentration of large grains enforces a 
decrease of $r_{\rm in}$ to keep a good match of the SED in the near-infrared. 
This, in turn,  leads to
a corresponding increase of temperature at the inner rim, 
$T_{\rm in}$, which is already very high (Tab.\ref{Tgraindist}). 
Thus, from the models with power-law grain-size distributions 
the best one is that with $\gamma= 3$. 

If the wind of \object{WR 118} is flattened due to some reason
(due to stellar rotation, binarity),
one can conclude from the circular symmetric shape of the two-dimensional 
visibility function,
that one looks face-on  at the flattened shell (``disk''). Williams et al. 
(1987) estimated the fraction of \object{WR\,118}'s  luminosity  absorbed by dust 
and re-radiated in the IR to be $\sim 71$\%. 
If the star is not obscured by its dust disk, we estimate that 
this fraction should not be less than 50\%.
This means that the dust disk should be optically 
thick and cover more than one half of the sphere.
Such a dust shell would not look like a 
geometrically thin disk,
but rather as a sphere with a hole in the light of sight, i.e. as  a
torus with a large opening angle.
However, first we should have clear indications of such asphericities, before we introduce 
a more complex geometry to the dust shell model of  \object{WR\,118}.
We note that van der Hucht (2001) suggested that 
all apparently single and persistently 
dust-making WC9 stars may possibly owe their
heated circumstellar dust signatures to colliding WC+OB wind effects.

Recently, Veen et al. (1998) have attributed eclipse-like variations of the
optical brightness of the dusty Wolf-Rayet stars
\object{WR\,103}, \object{WR\,121} and \object{WR\,113} 
to the obscuration of starlight by dust clouds in the
light-of-sight very close to the star, 
i.e. between the stellar surface and the permanent dust shell that is inferred from the IR excess. 
Moreover, they have estimated the sizes of the particles in the clouds to be of the order 
0.1\,$\mu$m. Crowther (1997) has discovered 
the visual fading of \object{WR\,104}, accompanied by the disappearance of high-ionization spectral 
lines, and placed the dust clouds inside the permanent dust shell. 
Thus, the decrease of the 
dust shell's inner boundary proposed by our modelling does not look as extraordinary as 
it appears with respect to the even more hostile conditions for grain condensation. 

\section{Conclusions}
We presented the first diffraction-limited 2.13\,$\mu$m speckle interferometric 
observations of the dusty Wolf-Rayet star \object{WR\,118} 
with 73 mas resolution. No evidence was found
for any significant deviation of the dust shell from circular symmetry. 
Radiative transfer calculations were carried out to model the spectral energy 
distribution and our 2.13\,$\mu$m visibility function.
The main conclusion of these calculations is that the grains 
in the permanent dust shell of the persistently dust-making Wolf-Rayet star
WR 118 grow to the relatively large sizes ($a \ge 0.3\,\mu$m).
They contribute more than 50\% in 
total mass of the dust shell and rises its 2.13\,$\mu$m albedo up to $\sim 0.35$.
The observations can be fitted either with a 
model containing double-sized grains  with radii of 
a = 0.05\,$\mu$m and 0.38$\mu$m or, alternatively, by a model with
a grain size distribution function $n(a) \sim a^{-3}$, with $a$ 
ranging between 0.005$\mu$m and 0.6$\mu$m.
At the inner boundary of the dust shell,
which has an angular diameter of $\Theta_{\rm in} = (17 \pm 1)$\,mas and
is located at $r_{\rm in} = (150 \pm 30)R_{\ast}$, 
the temperature of the smallest  grains  amounts to
1750\,K $\pm$ 100\,K and the dust-shell density to 
$(1 \pm 0.2) \cdot 10^{-19}$\,g/cm$^{3}$.
Adopting a wind velocity of $V_{\rm wind} = 1200$\,km/s, 
the dust formation rate is found to be $(1.3 \pm 0.5) \cdot 10^{-7}$ M$_{\odot}$/yr. 

The existence of large grains in the dust shell leads to a revision  
of the previously assumed size of the inner dust-shell boundary.
The inner dust-shell boundary is then located even closer to the star leading
to higher grain temperatures. 
This, in turn, further worsens the conditions of grain condensation which 
are already extremely harsh in any case. The question how  dust grains 
permanently condense under such conditions is still yet open  
(see Cherchneff et al.\ 2000 and references therein)
but the existence of dust shells
around Wolf-Rayet stars as in the case of \object{WR\,118} clearly
shows its warrant urgency.

%All these conclusions persist if the dust shell is flattened and observed almost face on.
%Thus, on the second glance,
%the decrease of the 
%dust shell's inner boundary proposed by our modelling does not look as extraordinary as 
%it appears with respect to the even more hostile conditions for grain condensation. 

\begin{acknowledgements}
The observations were made with  the SAO 6\,m telescope, operated by the
Special Astrophysical Observatory, Russia.
\end{acknowledgements}


\begin{thebibliography}{}
%
\bibitem{}
Allen D.A., Hyland A.R., Longmore A.J., Caswell J.L., Goss W.M., Haynes R.F., 1977, 
    ApJ 217, 108

\bibitem{}
Cherchneff I., Le Teuff Y.H., Williams P.M., Tielens A.G.G.M., 2000, A\&A 357, 572

\bibitem{}
Cohen M., Vogel S.T., 1978, MNRAS 185, 47

\bibitem{}
Crowther P.A., 1997, MNRAS 290, L59

\bibitem{}
Draine B.T., 1985, ApJS 57, 587

\bibitem{}
Egan M.P., Leung C.M., Spagna G.F., 1988, Comput. Phys. Comm. 48, 271

%\bibitem{}
%Hofmann K.-H., Weigelt G., 1986, A\&A 167, L15
%
\bibitem{}
IRAS Science Team, 1988, Explanatory Supplement to the IRAS Point Source 
   Catalogue, NASA

\bibitem{}
Ivezic Z., Elitzur M., 1997, MNRAS 287, 1997

\bibitem{}
J\"ager C., Mutschke H., Henning T., 1998, A\&A 332, 291

\bibitem{}
Labeyrie A., 1970, A\&A 6, 85

\bibitem{}
Leung C.M., 1976, ApJ 209, 75

%\bibitem{}
%Lohmann A.W., Weigelt G., Wirnitzer B., 1983, Appl. Opt. 22, 4028 
%
\bibitem{}
Massey P., Conti P.S., 1983, PASP 95, 440.

\bibitem{}
Roche P.F.,  Aitken D.A., 1984, MNRAS 208, 481

\bibitem{}
Schutte W.A., van der Hucht K.A., Whittet D.C.B., et al., 1998, A\&A 337, 261

\bibitem{}
van der Hucht K.A., 2001, New Astronomy Reviews 45, 135

\bibitem{}
van der Hucht K.A., Cassinelli J.P., Williams P.M., 1986, A\&A 168, 111

\bibitem{}
van der Hucht K., Williams P.M., Morris P.W., et al., 1996, A\&A 315, L193

\bibitem{}
Veen P.M., van Genderen A.M., van der Hucht K.A., Li A., Sterken A.,
Dominik C., 1998, A\&A 329, 199

%\bibitem{}
%Weigelt G., 1977, Optics Commun. 21, 55
%
%\bibitem{}
%Weigelt G., 1991, in: Progress in Optics Vol. 29, E. Wolf (ed), Elsevier Science 
%   Publishers, Amsterdam, p.293
%
\bibitem{}
Williams P.M., van der Hucht K.A., The P.S., 1987, A\&A 182, 91

\bibitem{}
Williams P.M., van der Hucht K.A., Morris P.W., 1998, Ap\&SS  255, 169

\bibitem{}
Williams P.M., van der Hucht K.A., 2000, MNRAS 314, 23


\end{thebibliography}
\end{document}